\documentclass[fleqn,usenatbib]{mnras}

\usepackage{newtxtext,newtxmath}
\usepackage{xcolor}

\usepackage[T1]{fontenc}

\DeclareRobustCommand{\VAN}[3]{#2}
\let\VANthebibliography\thebibliography
\def\thebibliography{\DeclareRobustCommand{\VAN}[3]{##3}\VANthebibliography}

\usepackage{graphicx}	
\usepackage{amsmath}	

\newcommand\musc{MUSCLES}
\newcommand{\Msun}{\mbox{$\mathrm{M}_{\odot}$}}
\newcommand{\Rsun}{\mbox{$\mathrm{R}_{\odot}$}}

\newcommand{\snr}{\ensuremath{S/N}}

\newcommand{\Teff}{\mbox{$T_{\mathrm{eff}}$}}
\newcommand{\logg}{\mbox{$\log g$}}
\newcommand{\pid}{17778}
\newcommand{\gj}{GJ\,207.1}
\newcommand{\gtwo}{G\,203-47}
\newcommand{\lhs}{LHS\,1817}
\newcommand{\wolf}{Wolf\,1130}
\newcommand{\hst}{\textit{HST}}
\newcommand{\gaia}{\textit{Gaia}}

\title[Revealing hidden WDs in PCEBs within 20 pc]{Direct detections of white dwarfs in four WD+dM post-common envelope binaries within 20 pc}

\author[M. W. O'Brien et al.]{
Mairi~W.~O'Brien$^{1}$\thanks{E-mail: mairi.obrien1@gmail.com},
David~J.~Wilson$^{2}$,
Pier-Emmanuel Tremblay$^{1}$,
Boris~T.~G\"ansicke$^{1}$,
Conor~M.~Byrne$^{1}$,
\newauthor Felipe~Lagos-Vilches$^{3}$,
J.~Sebastian~Pineda$^{2}$ 
and Joaqu\'in~A.~Barraza-Jorquera$^{3,4}$
\\
$^{1}$ Department of Physics, University of Warwick, Coventry CV4 7AL, UK \\
$^{2}$Laboratory for Atmospheric and Space Physics, University of Colorado, 600 UCB, Boulder, CO 80309, USA\\
$^{3}$Departamento de F\'isica, Universidad T\'ecnica Federico Santa Mar\'ia, Av. Espa$\tilde{n}$a 1680, Valpara\'iso, Chile\\
$^{4}$Instituto de F\'isica, Pontificia Universidad Cat\'olica de Valpara\'iso, Avenida Universidad 330, Valpara\'iso, Chile
}

\date{Accepted 2026 June 19. Received 2026 June 18; in original form 2026 May 22.}

\pubyear{\the\year{}}

\begin{document}
\label{firstpage}
\pagerange{\pageref{firstpage}--\pageref{lastpage}}
\maketitle

\begin{abstract}
Characterising post-common envelope binaries (PCEBs) containing a white dwarf and a main-sequence companion is essential for improving theories of binary evolution. This paper presents the first direct spectroscopic confirmations of the white dwarf components in four PCEB systems within 20\,pc of the Sun: \gtwo, \gj, \lhs, and \wolf. To detect the white dwarfs we obtained near-UV spectroscopy from STIS on the \textit{Hubble Space Telescope}, fitting with white dwarf models and M\,dwarf proxy spectra. We provide estimates of the white dwarf effective temperatures, which range from approximately 5300\,K to 6300\,K. We compare these parameters to those determined from modelling with photometry alone, and find a 5\,--\,8\,per\,cent discrepancy, due to emission features. Notably, 27 years after its initial detection, we confirm the presence of a white dwarf in \gtwo, which is the ninth closest white dwarf to the Sun. Using \textit{Swift} XRT data, we find that despite the 14.9-day orbital period of \gtwo, it is not tidally locked, possessing a rotation period likely exceeding 100\,days, and making it a rare example of a long-period PCEB formed via a brief common envelope interaction. We update the local white dwarf space density to (5.2\,$\pm\,$0.4)\,$\times$\,10$^{-3}$\,pc$^{-3}$, and compare our results to models from the Binary Populations and Spectral Synthesis (\textsc{bpass}) framework, finding a good agreement with the predicted and observed numbers of PCEBs within 20\,pc.
\end{abstract}




\section{Introduction}

White dwarfs in binaries with main-sequence companions, with separations less than $\approx$\,0.2\,au, are thought to form through common envelope (CE) evolution. CE evolution is a short-lived phase where both stars are enclosed in the envelope formed by the white dwarf progenitor as it undergoes post-main sequence evolution \citep{Paczynski1976}. The physics behind CE evolution is still poorly understood (see e.g. \citealt{Ivanova2013}). CE evolution must occur frequently due to the number of observed binaries containing a white dwarf with small separations, which can only be explained due to a loss of angular momentum during the CE phase \citep{Kruckow2021}. Detached post-common envelope binaries (PCEBs) are often the immediate product of CE evolution, and orbital angular momentum will be further lost in these systems until the main sequence star overflows its Roche lobe, and the system enters the mass transfer regime and eventually becomes a cataclysmic variable.

The orbital period distribution of PCEBs provides an observational constraint to support our understanding of CE evolution. Early large-scale studies of PCEBs, including the SDSS WD\,$+$\,MS binary catalogue \citep[e.g.][]{Rebassa2007,Rebassa2010,Rebassa2012a}, were sensitive to systems with short periods. The SDSS PCEB period distribution covered 2\,hours to 4\,days, with no indication of a substantial population in the days-to-weeks period regime \citep{Nebot2011}. A distinct population of wide-period PCEBs has emerged from both population synthesis and observational studies, with periods of hundreds to thousands of days \citep[e.g.][]{Escorza2019,Kruckow2021,Shahaf2023,Belloni2024}. This wide-period population contains more massive AFGK secondary stars, whereas the shorter-period population is dominated by M\,dwarf companions. Only a handful of PCEBs have been found which straddle the two populations, making this regime poorly constrained observationally.

Volume-limited, unbiased samples of white dwarfs enable detailed studies of the statistics and demographics of these stellar remnants. Since the white dwarf luminosity function peaks at low luminosities, the majority of local white dwarfs are relatively cool. The accurate parallaxes provided by the \textit{Gaia} spacecraft have enabled the creation of a volume-complete sample of white dwarfs within 40\,pc, which contains over a thousand stars, almost half of which are cooler than 6000\,K \citep{OBrien2024}. The only missing piece of this large and complete sample are the cool white dwarfs that are hidden in close binaries with bright main-sequence companions, as part of PCEBs. 

\begin{figure}
    \centering
	\includegraphics[width=\columnwidth]{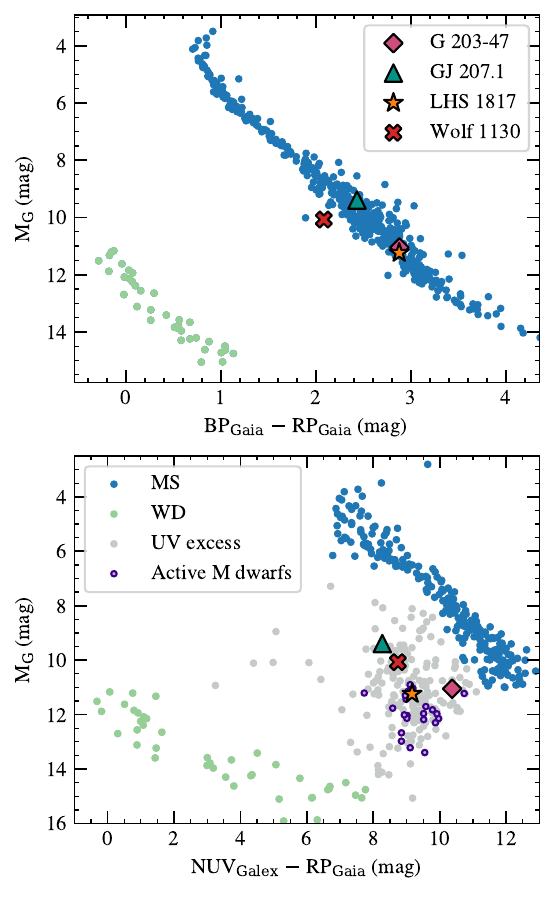}
	\caption{\textit{Top}: A colour--magnitude diagram showing \gaia\ absolute $G$ magnitude against \gaia\ $BP$ minus $RP$ magnitude for \gaia\ sources within 20\,pc of the Sun. \textit{Bottom}: A colour--magnitude diagram showing \gaia\ absolute $G$ magnitude against \textit{GALEX} $NUV$ minus \gaia\ $RP$ magnitude for 20\,pc \gaia\ sources. The four PCEBs discussed in this work are shown by the following symbols: \gtwo\ is the pink diamond, \gj\ is the teal triangle, \lhs\ is the orange star, and \wolf\ is the red cross. A sample of active M\,dwarfs from \citet{Pass2023} are shown as purple circles.}
    \label{fig:HR_UV}
\end{figure}

If the orbital alignment is favourable, PCEBs can show eclipses in their lightcurves (e.g. \citealt{Maxted2007}), however simulations from \citet{Parsons2013} identified that just 12\,per\,cent of PCEB systems eclipse. There are no known eclipsing PCEBs within 20\,pc of the Sun. If the white dwarf component of the PCEB is hot enough, it will demonstrate a clear UV photometric excess in surveys like \textit{GALEX} \citep{Parsons2016}. However, most white dwarfs are cool and faint even in the UV, and most M\,dwarfs are often active, causing photometric UV excesses that mimic a possible white dwarf companion. We show the location of some active M\,dwarfs from \citet{Pass2023} in a UV-optical colour-magnitude diagram in Fig.~\ref{fig:HR_UV}, noting that they have similar near-UV excesses to PCEBs. This combination of factors means that photometric UV excesses are not sufficient for directly detecting a white dwarf in a PCEB with an M\,dwarf. 

Cool white dwarf $+$ M\,dwarf (WD\,$+$\,dM) PCEBs in the local volume have been identified serendipitously via radial velocity (RV) surveys searching for exoplanets. Within 20\,pc, four of these systems have been discovered: \gtwo\ at 7.60\,pc \citep{Delfosse1999}, \gj\ at 15.78\,pc \citep{Baroch2021}, \lhs\ at 16.28\,pc \citep{Winters2020}, and \wolf\ at 16.58\,pc \citep{Gizis1998}. These systems are all M\,dwarfs or M subdwarfs with white dwarf-mass degenerate companions that are not detected in the optical. These systems are highlighted in Fig.~\ref{fig:HR_UV}, showing that they lie on the main sequence in a \gaia\ colour-magnitude diagram but demonstrate a near-UV excess. In this work, we present the first direct detections of the white dwarf components of these four PCEB systems, using \textit{Hubble Space Telescope} (\hst) and \textit{Swift} observations to confirm and characterise the white dwarfs. 

In Section~\ref{sec:properties} we describe the four PCEB systems that we analyse in this work, and in Section~\ref{sec:obs} we describe the observations used to detect the white dwarfs and the methods used to calibrate and fit those observations. Then, in Section~\ref{sec:results} we show our direct confirmations of the four white dwarfs in the PCEBs, estimating the effective temperatures of the white dwarfs. We use these new detections to revise the local white dwarf space density and the local PCEB space density. We then compare the numbers of observed PCEBs to those predicted by the Binary Populations and Spectral Synthesis (\textsc{bpass}) framework. Additionally, we analyse \gtwo, a system that sits between two distinct populations of PCEBs, comparing its orbital and rotation periods to other similar PCEBs. We conclude in Section~\ref{sec:conclusion}.

\section{Sample properties}
\label{sec:properties}

The four PCEBs within 20\,pc that we analysed in this work were identified due to their RV shifts. However, there are additional indicators that the four M\,dwarfs have hidden white dwarf companions. Fig.~\ref{fig:HR_UV} shows that all four PCEBs display excesses in \textit{GALEX} $NUV$, however many other M\,dwarfs in 20\,pc also demonstrate this excess due to activity \citep{Pass2023}. Additionally, three of our systems have a high \gaia\ renormalised unit weight error (RUWE), which is a possible indicator of multiplicity. For a single-star system, RUWE is expected to be $\lesssim$\, 1.4. \gtwo\ has a RUWE of 23.2, \gj\ has a RUWE of 2.43, and \wolf\ has a RUWE of 1.82. However, \lhs\ has a RUWE of 1.36, making it a borderline case for multiplicity. There are many overlapping indicators of close white dwarf companions to M\,dwarfs, but the only way to confirm these companions directly is with near-UV spectroscopy.

Each individual system is discussed below:

\textbf{\gtwo}: \gtwo\ was identified as a single-lined spectroscopic binary by \citet{Reid1997}. Assuming an M\,dwarf mass of 0.2\,\Msun, the mass of the companion star in the system was determined by \citet{Delfosse1999} using HIPPARCOS data \citep{Hipparcos1997}, to be $\gtrsim$\,0.5\,\Msun, implying that the companion star must be degenerate. The most up-to-date orbital period of the system determined by \citet{Shariat2026} is 14.909777\,$\pm$\,0.00001\,days. At a \gaia\ DR3 distance of 7.599\,$\pm$\,0.025\,pc, the white dwarf in \gtwo\ is the ninth closest white dwarf to the Sun. Despite its accurate parameters, proximity, and \textit{GALEX} excess, it has taken over two decades following its RV identification to obtain a UV spectrum of this system. 

\textbf{\gj}: \gj\ is commonly known as Wachmann's Flare Star. \gj\ was observed as part of the CARMENES survey searching for exoplanets orbiting M\,dwarfs \citep{Quirrenbach2018}, and was flagged as a single-lined spectroscopic binary by \citet{Baroch2021}. With the M\,dwarf mass assumed to be 0.49\,\Msun, \citet{Baroch2021} determined a degenerate companion mass of 0.63\,$^{+0.20}_{-0.13}$\,\Msun. \citet{Bar2017} observed \gj\ with the X-shooter spectrograph on the VLT and placed an upper limit on the \Teff\ of the white dwarf at 8000\,K, due to their non-detection. The orbital period of the system determined by \citet{Baroch2021} is 0.60417356$^{+0.00000024}_{-0.00000026}$\,days. The \gaia\ DR3 distance to \gj\ is 15.783\,$\pm$\,0.012\,pc.

\textbf{\lhs}: \lhs\ was identified as a single-lined binary by \citet{Winters2020}. They determined the mass of the degenerate companion using two different methods of estimating the orbital inclination of the system, giving masses of 1.03\,$\pm$\,0.08\,\Msun\ and 0.64\,$^{+0.25}_{-0.14}$\,\Msun. For our analysis we assume that the mass range is between the minimum of the smaller of these mass measurements up to the maximum of the larger mass measurement. The orbital period of the system determined by \citet{Winters2020} is 0.30992678\,$\pm$\,0.00000048\,days. The \gaia\ DR3 distance to \lhs\ is 16.280\,$\pm$\,0.014\,pc.

\textbf{\wolf}: A degenerate companion to the M subdwarf in \wolf\ was identified by \citet{Gizis1998}. A T8 subdwarf was found to be a wide tertiary companion to \wolf\ by \citet{Mace2013}. \citet{Mace2018} determined an orbital period for the WD\,$+$\,dM system of 0.4966\,$\pm$\,0.0001\,days, and a white dwarf companion mass of 1.24$^{+0.19}_{-0.15}$\,\Msun. Using \textit{TESS} data, \citet{Honaker2020} also determined that the degenerate companion is likely to be a highly massive white dwarf. \citet{Mace2018} reported a non-detection of a white dwarf in their analysis of a UV spectrum from \hst. The \gaia\ DR3 distance to \wolf\ is 16.585\,$\pm$\,0.007\,pc.

\section{Observations and Methods}
\label{sec:obs}
\subsection{\textit{HST}}
\label{sec:hstobs}

\begin{table*}
    \centering
    \caption{Details of the four PCEBs and their \hst\ STIS observations that were analysed in this work. Distances were derived from \gaia\ DR3 parallaxes, and orbital periods are from the following references: a) \citet{Shariat2026}, b) \citet{Baroch2021}, c) \citet{Winters2020}, d) \citet{Mace2018}.}
    \begin{tabular}{lllllll}
    \hline 
    Name &   Distance&Orbital period&UV grating and aperture& Optical grating&  Program&Date of \\
    &   (pc)&(days)&& and aperture&  ID&observation\\ \hline
    \gtwo&   7.599\,$\pm$\,0.025&14.909777\,$\pm$\,0.00001\textsuperscript{a}&G230L 52\,$\times$\,2; G230LB 52\,$\times$\,0.2& -- & 17778& 2026 Mar 25\\
    \gj&   15.783\,$\pm$\,0.012&0.60417356$^{+0.00000024}_{-0.00000026}$\textsuperscript{b}&G230LB 52\,$\times$\,0.2& G430L 52\,$\times$\,2&  17778&2025 Aug 13 \\
    \lhs&   16.280\,$\pm$\,0.014&0.30992678\,$\pm$\,0.00000048\textsuperscript{c}&G230LB 52\,$\times$\,0.2& G430L 52\,$\times$\,2& 17778&2025 Sept 24 \\
    \wolf&   16.585\,$\pm$\,0.007&0.4966\,$\pm$\,0.0001\textsuperscript{d}&G230LB 52\,$\times$\,0.2& G430L 52\,$\times$\,0.2& 9786&2003 Aug 10 \\
     \hline
    \end{tabular}
    \label{tab:hst_obs}
\end{table*}

The \hst\ observations of \gj, \lhs\ and \gtwo\ were carried out during Cycle 32 under PID\,17778. All observations were made using the Space Telescope Imaging Spectrograph (STIS; \citealt{Woodgate1998}). \gtwo\ was observed in the ultraviolet with the NUV-MAMA G230L grating, due to its low activity levels. The G230L grating is low-resolution with R\,$\sim$\,500, and well flux-calibrated. \gj\ and \lhs\ contain active M\,dwarfs that flare regularly, and if the systems were to be observed during a flare state, they would surpass the strict NUV-MAMA bright object limits. Therefore, observations of these systems with the G230L grating were prohibited due to instrument safety constraints. The R\,$\sim$\,700 G230LB grating, which uses the STIS CCD, has no bright object limits so was used instead for the \gj\ and \lhs\ observations. \gtwo\ was observed with G230LB as well as G230L for a direct comparison between the gratings.

\gj\ and \lhs\ were observed in the optical with the STIS CCD G430L grating, which is a low R\,$\sim$\,500 grating covering 2900\,--\,5700\,\AA. The combination of G230L/G230LB and G430L provides a good \snr\ between 2200\,--5700\,\AA. The details of these observations are shown in Table~\ref{tab:hst_obs}. An error was made in the setup of the G430L observation of \gtwo, and therefore the exposure was too short to be useful. Because of this, we incorporated the \gaia\ XP spectrum of \gtwo\ into our fit \citep{DeAngeli2023}. To improve the flux calibration of the \gaia\ XP spectrum, we applied the corrections derived by \citet{Huang2024}. 

In addition to the three systems observed in PID\,17778, \wolf\ has existing STIS observations published in \citet{Mace2018}. \wolf\ was observed on 2003--08--10 with STIS G230LB, G430L and G750L as part of PID\,9786, the details of which are shown in Table~\ref{tab:hst_obs}. These observations were included in the Low Resolution Stellar Library (LOWLIB) High Level Science Product \citep{Pal2023}\footnote{\url{https://archive.stsci.edu/hlsp/lowlib}}. However, the G230LB spectrum was not properly extracted for the version included in LOWLIB. We re-extracted the spectra from the crj files using {\sc stistools}\footnote{\url{https://stistools.readthedocs.io/en/latest/}}. We then co-added the spectra using the Hubble Advanced Spectral Products (HASP) routines, with the data quality flags (DQ) set to zero, so that the HASP routines would accept the spectra. The co-added spectrum was then appended to the LOWLIB G430L spectrum.  

\begin{figure}
    \centering
    \includegraphics[width=\linewidth]{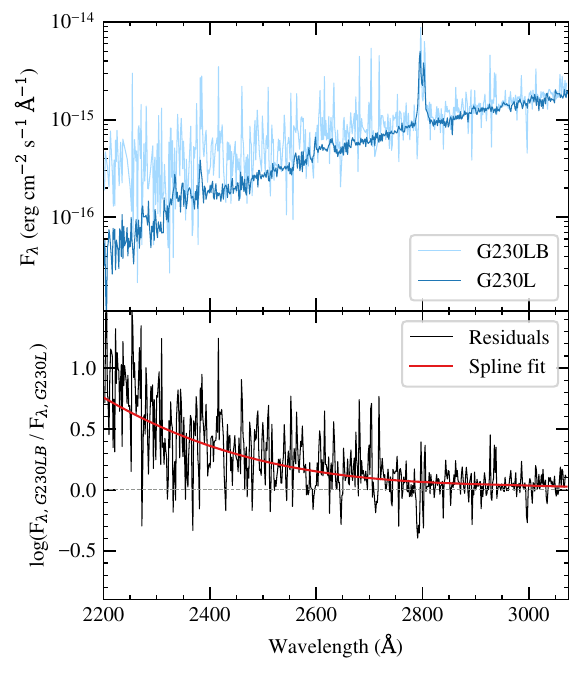}
    \caption{STIS G230L (dark blue) and G230LB (light blue) spectra of \gtwo, with residuals (black) showing the difference between the fluxes from both gratings on a log scale. Noise features from the G230LB spectrum have been masked. The red line shows the spline fit to the residuals, which is then used as a correction function for the other G230LB spectra in this work.}
    \label{fig:G230L_vs_G230LB_with_residuals}
\end{figure}

The G230LB grating scatters red light into the CCD, and for red objects the scattered light can make a significant contribution to the detected ``ultraviolet'' flux and perhaps mimic the appearance of a blue companion star. We tried correcting the G230LB spectra following the standardised recipe of \citet{Worthey2022}\footnote{\url{https://hst-docs.stsci.edu/stisihb/chapter-13-spectroscopic-reference-material/13-3-gratings/first-order-grating-g230lb}}. However, we found that the \citet{Worthey2022} correction was not sufficient for our spectra, as their model assumes the red star is the sole source of both the signal and the scattering, whereas in our case the white dwarfs contributed substantially to the G230LB flux. For a direct comparison between the flux calibration of G230L and G230LB gratings for our PCEBs, we observed \gtwo\ with G230L for one orbit and with G230LB for another orbit, which we compare in the top panel of Fig.~\ref{fig:G230L_vs_G230LB_with_residuals}. We derived a wavelength-dependent correction function by fitting a smooth spline function to the logarithmic flux ratio of the \gtwo\ G230LB and G230L spectra, in the wavelength region between 2200\,\AA\ and 3075\,\AA. We calculated the log-ratio of the flux densities, $\log_{10}(F_{\lambda, \text{G230LB}} / F_{\lambda, \text{G230L}})$, and fitted the resulting residuals with a cubic smoothing spline, $S(\lambda)$, which is shown as the red line in Fig.~\ref{fig:G230L_vs_G230LB_with_residuals}. The resulting function that can be applied to other G230LB spectra is,
\begin{equation}
    F_{\lambda, \text{G230LB, corr}} = F_{\lambda, \text{G230LB}} \times 10^{-S(\lambda)}.
\end{equation} 

\noindent This function was then applied to the other three G230LB spectra to reduce the impact of the red leak.

\begin{table*}
    \centering
        \caption{Available Swift datasets for our targets. Exposure times are approximate as the different instruments have slightly different recorded exposure times. Datasets marked Y in the final two columns were used to build the X-ray spectra and obtained simultaneously with the HST spectra respectively.}
    \begin{tabular}{llllllll}
    \hline
    Name & Observation ID & Date & Exposure time& UVOT filter & UVM2 flux & Used XRT? & HST? \\
    &&&(s)&& ($10^{-16}$ erg s$^{-1}$ cm$^{-2}$ \AA$^{-1}$)&& \\ \hline
\gtwo & 15937001 & 2023 Mar 31& 955 & UVM2 & $2.1\pm0.2$ & Y & -- \\
 & 15937002 & 2023 Mar 31& 293 & U & -- & Y & -- \\
 & 15937003 & 2023 Apr 02& 762 & UVM2 & $2.1\pm0.2$ & Y & -- \\
 \gj & 14457001 & 2021 Aug 07& 906 & UVW2 & -- & -- & -- \\
 & 14457002 & 2021 Nov 13& 506 & UVM2 & $12.6\pm0.4$ & Y & Y \\
 & 14457003 & 2025 Aug 13& 2915 & UVW2 & -- & -- &-- \\
\lhs & 36027001 & 2006 Aug 19& 2276 & UVM2 & $20.4\pm0.5$ & Y & -- \\
 & 36027002 & 2025 Sept 24& 955 & UVM2 & $14.9\pm0.5$ & Y & Y \\
\wolf & 40011002 & 2011 Jul 27& 1163 & UVW2 & -- & -- & -- \\
 & 40011003 & 2024 Jan 10& 1361 & UVM2 & $4.8\pm0.3$ & Y & -- \\
 & 40011004 & 2025 Sept 21& 359 & UVM2 & $4.4\pm0.5$ & -- & -- \\
 \hline
    \end{tabular}
    \label{tab:swiftobs}
\end{table*}

\subsection{\textit{Swift}}

The ultraviolet continuum enhancement during a strong flare could mimic both a white dwarf spectrum and the G230LB red scatter, so it is important to rule out or remove any flares from our observations. For \gtwo, we extracted a light curve from the time-tagged G230L spectrum and found no evidence for flares. As the STIS G230LB grating uses the non-photon counting CCDs, we were unable to extract light curves from the data for the other systems to search for and remove flares. We therefore obtained simultaneous observations with the Neil Gehrels Swift Observatory (\textit{Swift}), using the UVM2 filter on the Ultraviolet/Optical Telescope (UVOT) in event mode to track the activity of the targets during the \hst\ observations. 

\gj\ was observed with \textit{Swift} for two orbits, almost exactly matching the \hst\ time on target. \lhs\ was observed with \textit{Swift} for the last quarter of its \hst\ observation. Simultaneous observations were not possible for \gtwo\ because \textit{Swift} had moved into a limited observing mode during the observation window, in preparation for its attempted re-boost. Archival UVM2 observations are available for both \gtwo\ and \wolf. Details of the \textit{Swift} datasets used are given in Table \ref{tab:swiftobs}. UVOT data was analysed with the relevant {\sc HEASOFT}\footnote{\url{https://heasarc.gsfc.nasa.gov/docs/software/lheasoft/}} packages. UVM2 light curves were extracted at 10\,s and 100\,s cadence from the event files and fluxes were measured from the final image products. \textit{Swift} X-ray Telescope (XRT) light curves and spectra were extracted using the online XRT tool\footnote{\url{https://www.swift.ac.uk/user_objects/}}. For the two simultaneous observations, the UVOT light curves showed no evidence for significant flares, so we assume that the \hst\ spectra represents the stars in a ``normal'' activity state. For \wolf, the two G230LB spectra are identical within their uncertainties. The blue end of the G230LB spectrum of \gtwo\ (i.e. the region less affected by scattered light) shows no significant differences from the G230L spectrum. In both cases, a flare in one or both spectra would produce flux differences. We therefore confidently rule out significant NUV flux contributions from flares in all of our spectra. 

We initially planned to use the UVM2 data to assess the red leak corrections described in Section \ref{sec:hstobs}, integrating the \hst\ observations over the UVM2 bandpass to compare the fluxes. However, we found that for three out of four cases the \textit{Swift} flux was higher than even the uncorrected G230LB spectrum. We suspect this is a combination of the low \snr\ of the STIS data in the bandpass and incomplete characterisation of red leak in the UVM2 filter. Whatever the cause, we elected not to incorporate the \textit{Swift} data into the red leak removal process or into our spectroscopic fits.

\subsection{Spectroscopic fitting}
\label{sec:fitting}

\begin{table*}
    \centering
        \caption{Stellar parameters and measured X-ray fluxes for the M\,dwarf components of the PCEBs, and stellar parameters of the white dwarf components determined in this work. References correspond to the origins of the M\,dwarf parameters as well as white dwarf masses.}
    \begin{tabular}{llllllllll}
        \hline 
        Name &  $M_\mathrm{MS}$ &\Teff$_\mathrm{,MS}$ & $R_\mathrm{MS}$ & $L_\mathrm{MS}$ & $L_\mathrm{x}$ ($10^{28}$&  $M_\mathrm{WD}$& \logg$_\mathrm{WD}$  &\Teff$_\mathrm{,WD}$ &References  \\
 &  (\Msun)&(K)& (\Rsun)& ($\mathrm{L}_{\odot}$)& erg s$^{-1}$)&  (\Msun)& (fixed) &(K) &\\   \hline
        \gtwo&  $0.27\pm0.01$&$3279^{+65}_{-63}$& $0.29\pm0.01$& $0.0085\pm0.0001$& $<0.1$ &  $\gtrsim$\,0.5 & 8.0\,$^{+1.00}_{-0.13}$ &5286\,$^{+788}_{-91}$ &\citet{Delfosse1999};\\
 &  && & & & & & &\citet{Pineda2021}\\
        \gj&  $0.49\pm0.02$&$3419\pm51$& $0.48\pm0.01$& $0.029\pm0.001$& $9.18^{+0.74}_{-0.46}$ &  0.5 -- 0.83 & 8.0\,$^{+0.37}_{-0.13}$ &6100\,$^{+400}_{-150}$ &\citet{Baroch2021}\\
        \lhs&  $0.27\pm0.01$&$3220^{+69}_{-65}$& $0.29\pm0.01$& $0.008\pm0.0002$& $2.50^{+0.23}_{-0.26}$ &  0.5 -- 1.11& 8.0\,$^{+0.80}_{-0.13}$ &6300\,$^{+900}_{-100}$ &\citet{Winters2020}\\
        \wolf&  $0.30\pm0.01$&$3530\pm60$& $0.29\pm0.01$& $0.012\pm0.00$& $3.14^{+0.52}_{-0.55}$ &  1.09 -- 1.43 & 8.8\,$^{+0.20}_{-0.01}$ &6100\,$^{+100}_{-100}$ &\citet{Mace2018}\\
        \hline
    \end{tabular}
    \label{tab:fitting_params}
\end{table*}

We fitted the combined and red leak-corrected STIS G230L/G230LB and G430L spectra of the four PCEB systems, using a combined model consisting of an M\,dwarf proxy, plus white dwarf atmosphere models. Stellar parameters for the M\,dwarfs were compiled from the literature and are shown in Table \ref{tab:fitting_params}. For \lhs\ we were unable to find a complete set of modern (i.e. post-\gaia) parameters, so we calculated a new set using the methods described in \citet{Pineda2021}. We used the parameters to select appropriate proxy stars with available high-quality NUV and optical spectra to model the M\,dwarf component, first by matching \Teff, then where possible finding a close match to the rotation period. For \gtwo, \gj\ and \lhs\ we used Barnard's Star, GJ\,674 and GJ\,729 respectively, which all have spectra from the \musc\ survey \citep{Loyd2016, Wilson2025}. For Wolf 1130 we used the LOWLIB spectrum of Kapteyn's Star \citep{Pal2023}. White dwarfs typically have either hydrogen or helium-rich atmospheres, and given that the white dwarfs in the PCEBs have been accreting wind from their M\,dwarf companions for many Gyr, we assumed their atmospheres are hydrogen-rich. We therefore used updated 3D local thermodynamic equilibrium (LTE) model spectra\footnote{\url{https://warwick.ac.uk/fac/sci/physics/research/astro/people/tremblay/modelgrids/}} for hydrogen-atmosphere white dwarfs from \citet{Tremblay2013, Tremblay2015} to model the white dwarf components. 

For fitting, the Pan-STARRS or SDSS $r$-band magnitude of the M\,dwarf proxy spectrum was first scaled to the $r$-band magnitude of the PCEB, because the region redward of 5000\,\AA\ does not contain any substantial white dwarf flux. We ensured we were fitting just the white dwarf continuum by masking any emission features in the UV spectra, including Mg\,\textsc{ii}\,2800\,\AA. We combined the M\,dwarf proxy spectrum with the grid of hydrogen-atmosphere white dwarf models, and applied a reduced $\chi^2$ least-squares fitting method to return the best solution for the white dwarf effective temperature (\Teff), while fixing the surface gravity (\logg). \Teff\ is partially degenerate with the assumed \logg, and allowing \logg\ to vary within the range corresponding to the white dwarf RV mass introduces an additional source of uncertainty. We fixed \logg\ at 8, the canonical value for single white dwarfs \citep{Hollands2018_Gaia,Bergeron2019,Mccleery2020,OBrien2024}, for \gtwo, \gj\ and \lhs. \wolf\ has an RV mass constraining it to be much heavier (1.09\,--\,1.43\,\Msun), so we instead fixed \logg\ at 8.8. For the uncertainties on the white dwarf \Teff\ in Table~\ref{tab:fitting_params}, we fixed \logg\ to correspond to the lower and upper bounds of the masses determined from RV measurements, and then calculated the best-fitting \Teff\ for those \logg\ values.

\subsection{Photometric modelling}
\label{sec:phot}

To demonstrate the benefits of obtaining STIS near-UV spectra for PCEBs, we predict the white dwarf \Teff\ for our systems based on photometry alone. All four systems have \textit{GALEX} $NUV$ photometry \citep{GALEX2005}, and we applied the non-linearity correction of \citet{Wall2019} to the \textit{GALEX} $NUV$ photometry. In the optical, \gtwo\ and \gj\ have SDSS $ugriz$ photometry, and \lhs\ and \wolf\ have Pan-STARRS $grizy$ photometry available \citep{SDSS_dr16,PanSTARRS2016}. Due to the brightness of our targets, there were issues with saturation in some of the SDSS and Pan-STARRS bands, so we carefully selected uncontaminated data to the best of our ability. 

For our comparison, we used the typical setup of combining M\,dwarf models with white dwarf models, creating synthetic photometry, and comparing it to the observed photometry. This comparison also highlights the differences of using M\,dwarf templates compared to M\,dwarf models, as we used M\,dwarf templates in our spectroscopic fit. For this comparison we used \textsc{BT-Settl} M\,dwarf models \citep{Allard2012}, and pure-H 3D LTE white dwarf models \citep{Tremblay2013, Tremblay2015}. For \gtwo, \gj, and \lhs\ we used Solar metallicity \textsc{BT-Settl-CFIST} models, and for the sub-Solar metallicity M\,subdwarf \wolf\ we used \textsc{BT-Settl-AGSS2009} models at [Fe/H]\,$=$\,$-$0.7 \citep{Mace2018}. The models were scaled using the \gaia\ DR3 distances of the targets and the M\,dwarf models were generated using the effective temperatures, masses and radii listed in Table~\ref{tab:fitting_params}. The \logg\ for the white dwarf models were fixed at the same values as for the spectroscopic fit (\logg\,$=$\,8 for \gtwo, \gj\ and \lhs, and \logg\,$=$\,8.8 for \wolf). The \Teff\ of the white dwarf models was then varied to match the \textit{GALEX} $NUV$ photometry.

\section{Results and Discussion}
\label{sec:results}

\subsection{White dwarf parameters}
\label{sec:wd_teff}

\begin{figure*}
    \centering
    \includegraphics[width=\textwidth]{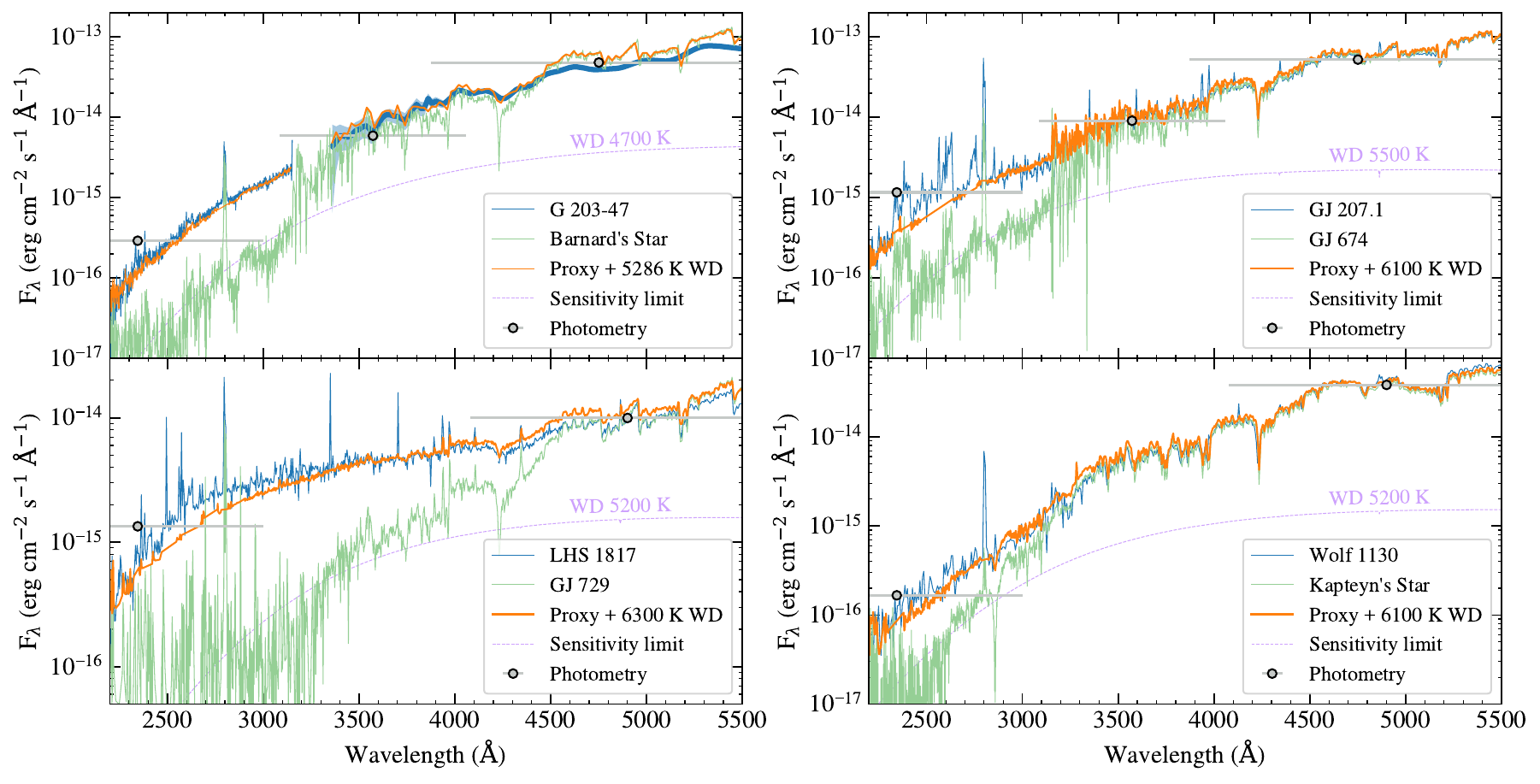}
    \caption{\hst\ STIS spectra (plus a \gaia\ XP spectrum for \gtwo) of the four PCEB systems (blue) compared with observations of M\,dwarf proxy stars (green) scaled to the PCEB M\,dwarf flux on the Pan-STARRS or SDSS $r$-band magnitude. The fit to the STIS spectra with a combined M\,dwarf proxy $+$ white dwarf model \citep{Tremblay2013,Wilson2025} is shown in orange, where emission features have been masked. Photometric data is shown in grey, but was not incorporated into the fit: GALEX $NUV$ (all four), SDSS $u$ and $g$ (\gtwo\ and \gj) and Pan-STARRS $g$ (\lhs\ and \wolf). The G230LB spectra have been binned to reduce noise. The purple dashed lines represent the sensitivity limit, i.e. the highest \Teff\ that a white dwarf could go undetected in each system.}
    \label{fig:all_four_stis_fits}
\end{figure*}

The best fitting M\,dwarf proxy plus white dwarf models to each PCEB spectrum are shown in Fig.~\ref{fig:all_four_stis_fits}. Due to the excellent flux calibration of the G230L spectrum of \gtwo\ we were able to fit the white dwarf component of the system without any corrections. With the white dwarf \logg\ fixed at 8, we determined \Teff\,$=$\,5286\,K for the white dwarf. The other three systems only have G230LB spectra available, which are noisy and less well flux-calibrated (see Fig.~\ref{fig:all_four_stis_fits}). Therefore, despite correcting for the G230LB red leak, we only provide estimates on the white dwarf temperatures of these systems using the available data. For \gj, we fixed \logg\ at 8, and found \Teff\,$\approx$\,6100\,K. For \lhs\ we also fixed \logg\ at 8, and found \Teff\,$\approx$\,6300\,K. For \wolf\ we fixed \logg\ at 8.8 due to the high RV mass, and found \Teff\,$\approx$\,6100\,K. These parameters are listed in Table~\ref{tab:fitting_params}. The errors reported in Table~\ref{tab:fitting_params} correspond to the \Teff\ values that were calculated with \logg\ values corresponding to the lower and upper bounds on the RV masses from Table~\ref{tab:fitting_params}. The maximum \logg\ in the \citet{Tremblay2013} grids is 9, and therefore we fitted up to a \logg\ of 9 for \gtwo\ and \wolf, which have upper mass limits above that corresponding to \logg\,$=$\,9.

Additionally, we calculated the highest \Teff\ of a white dwarf companion that would not be detected with \hst\ STIS G230L or G230LB for each system. Models providing these limits are shown in Fig.~\ref{fig:all_four_stis_fits}. For \gtwo, \gj, \lhs\ and \wolf\ we found \Teff\ limits of 4700\,K, 5500\,K, 5200\,K and 5200\,K respectively. The \logg\ values for these models were fixed at the same values as in the fit (all at 8 except \wolf\ which is at 8.8). This detection sensitivity test shows that the coolest white dwarfs in nearby PCEBs may not be detectable even with a high-quality, flux calibrated UV spectrum.

\subsection{Comparison with photometric SED modelling}
\label{sec:sed_fits}

\begin{figure*}
    \centering
    \includegraphics[width=\textwidth]{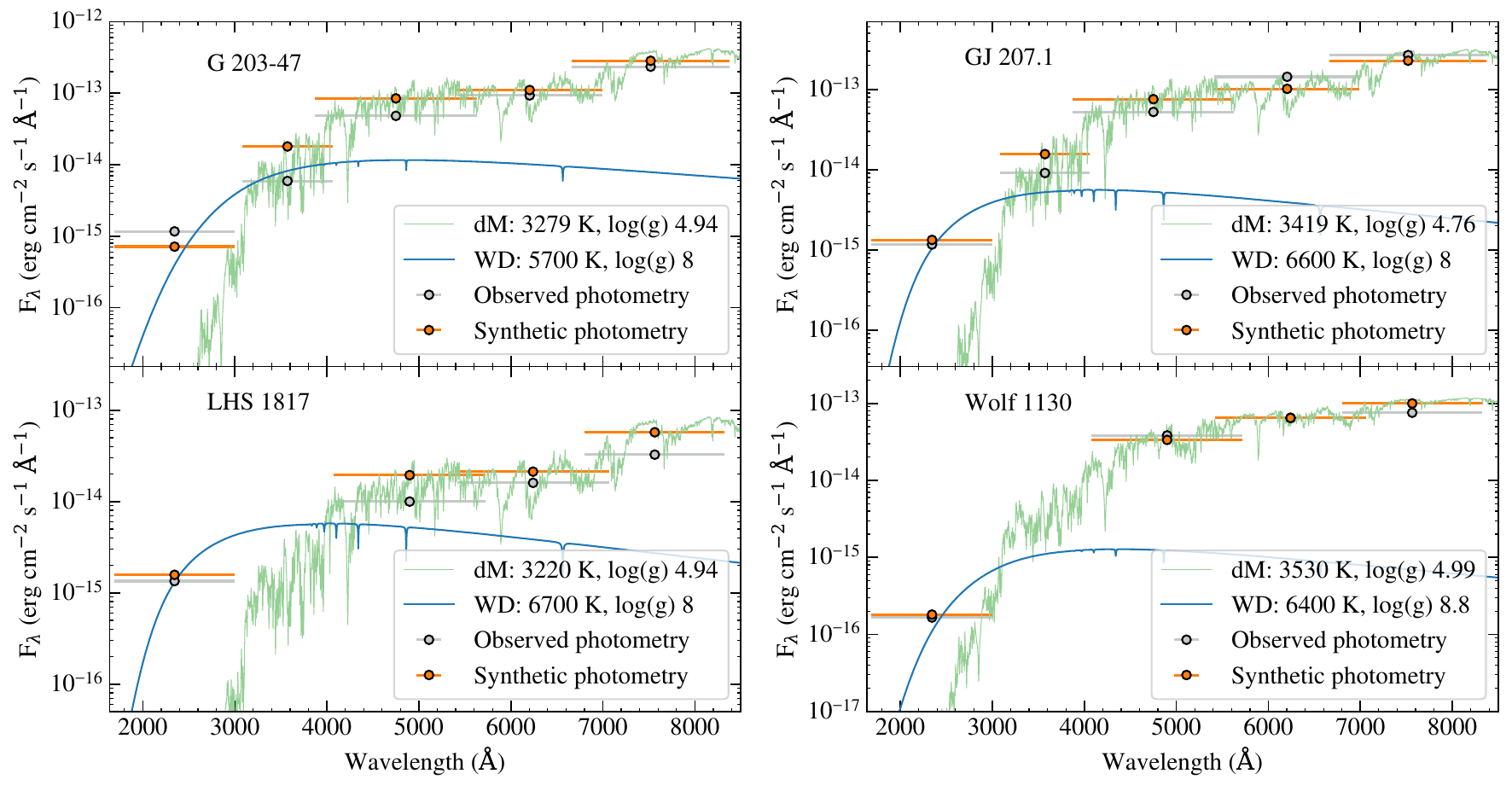}
    \caption{Modelling of the four PCEBs using M\,dwarf models from \textsc{BT-Settl-CIFIST} and \textsc{BT-Settl-AGSS2009} in green and white dwarf models in blue. The M\,dwarf models were calculated using parameters from Table~\ref{tab:fitting_params}. Photometric data is shown in grey: GALEX $NUV$ (all four), SDSS $ugri$ (\gtwo\ and \gj) and Pan-STARRS $gri$ (\lhs\ and \wolf). Synthetic photometry, calculated by combining the M\,dwarf model and white dwarf model, is shown in orange.}
    \label{fig:photometry_BT_Settl}
\end{figure*}

To quantify the benefits of observing PCEBs in the UV with STIS, we compare our spectroscopic fits with photometric modelling. Typically, M\,dwarf models and white dwarf models are combined to fit the photometry of a PCEB system, which we demonstrate in Fig~\ref{fig:photometry_BT_Settl}. Comparing Figs.~\ref{fig:all_four_stis_fits} and~\ref{fig:photometry_BT_Settl}, we see that the \Teff\ of the white dwarfs determined from modelling the photometry alone, and using M\,dwarf models, are 5\,--\,8\,per\,cent larger than from our STIS fits with M\,dwarf proxy spectra. This is in part due to the activity in some of the systems causing emission features in the \textit{GALEX} $NUV$ passband. The active systems \lhs\ and \gj\ are particularly affected by this issue, which we mitigated in the STIS fits in Fig.~\ref{fig:all_four_stis_fits} by masking the emission features and just fitting the continuum. Additionally, the M\,dwarf models are substantially fainter in the UV than the proxy spectra we chose for our STIS fits because the models do not include the chromospheric contributions that define the M\,dwarf UV spectra. Because of this, more flux contribution from a white dwarf is required to match the \textit{GALEX} $NUV$ photometry. We performed this test to provide a word of caution for fitting UV photometry of PCEBs without STIS spectra. We advise using M\,dwarf proxy spectra with a similar \Teff\ and activity level to the M\,dwarf, and obtaining flux-calibrated STIS spectra where possible, in order to distinguish between the white dwarf continuum and any emission features.

\subsection{Space density}
\label{sec:spacedensity}

\subsubsection{White dwarf space density}
\label{sec:WDspacedensity}

Following the confirmation of the white dwarfs in the four PCEBs discussed in this work, we calculated an up-to-date white dwarf space density using the 20\,pc volume. Despite our volume of interest being so small, in order to accurately calculate the space density we must consider the non-uniform Galactic distribution of stars. We assumed an old thin disc scale height of $h$\,$=$\,300\,pc \citep{Gilmore1983, Harris2006}, and determined the volume $V$ based on a vertically exponential distribution where the Sun's vertical distance is offset by 20\,pc from the Galactic midplane:
\begin{equation}
    V(z_{\rm max}) = 2 \pi \int_{0}^{z_{\rm max}} e^{-\frac{z+20}{h}} \left(z_{\rm max}^2 - z^2\right) \, dz,
\end{equation}

{\noindent}where $z$ is the distance measured from the Galactic plane in pc, $h$ is the 300\,pc scale height, and $z_{\rm max}$ is our radius of interest, 20\,pc. The resulting volume is 3.058\,$\times$\,10$^{4}$\,pc$^{3}$. 

The number of white dwarfs in the local 20\,pc volume includes all white dwarfs recognised in the \gaia\ DR3 white dwarf sample \citep{Hollands2018_Gaia,Gentile2021,OBrien2024}, as well as Procyon B, HD\,13445\,B, WD\,0727$+$482\,A$+$B, and the four PCEB systems discussed in this work (\gtwo, \gj, \lhs, \wolf). We count the confirmed double-lined system WD\,0135--052 as two white dwarfs. There are nine candidate unresolved double white dwarfs which have masses less than 0.53\,\Msun\ after applying the \citet{OBrien2024} correction for model-dependent mass underestimation. There is also a tentative detection of an ultra-cool white dwarf candidate found in \gaia\ DR3 by \citet{Golovin2024} (\gaia\ DR3 6013647666939138688). In total, we identified $N$\textsubscript{tot}\,$=$\,153 white dwarfs, plus up to 10 additional candidate white dwarfs or double degenerates that are not directly confirmed.

Following \citet{Hollands2018_Gaia}, we treated the observed white dwarf count as a Poisson realisation of an underlying mean $\bar{N}$, and inferred $\bar{N}$ using a Gamma posterior with a Jeffreys prior. We additionally treated the candidate unresolved double degenerates and other white dwarf candidates as part of a binomial distribution with $p\,=\,$0.5. The resulting $\bar{N}$\,$=$\,158.2\,$^{+12.8}_{-12.3}$ was then divided by the volume to calculate a local observed white dwarf space density of (5.2\,$\pm\,$0.4)\,$\times$\,10$^{-3}$\,pc$^{-3}$. This value of the white dwarf space density is a strict lower limit, as there are possibly more unidentified PCEBs within 20\,pc (see Section~\ref{sec:PCEBspacedensity} for a discussion).

\citet{Hollands2018_Gaia} calculated a white dwarf space density using the 20\,pc sample from \gaia\ DR2 of (4.49\,$\pm\,$0.38)\,$\times$\,10$^{-3}$\,pc$^{-3}$. Since the publication of \citet{Hollands2018_Gaia} and the release of \gaia\ DR3, both \lhs\ and \gj\ were identified as PCEBs \citep{Winters2020,Baroch2021}. The inclusion of \lhs, \gj\ and \gaia\ DR3 6013647666939138688, plus our treatment of double degenerate candidates, increased the 20\,pc space density by 16\,per\,cent compared to the \gaia\ DR2 value from \citet{Hollands2018_Gaia}. \citet{Munn2017} determined an independent measure of the disc white dwarf space density, (5.5\,$\pm\,$0.1)\,$\times$\,10$^{-3}$\,pc$^{-3}$, using the SDSS footprint. Their result agrees with our local space density within 1\,$\sigma$.

\subsubsection{PCEB space density}
\label{sec:PCEBspacedensity}

Using the same method as above, considering the four PCEB systems discussed in this work, we determined an observed WD\,$+$\,dM PCEB space density. The mean of the Poisson distribution is $\bar{N}$\,$=$\,4.2\,$^{+2.4}_{-1.7}$ and the resulting PCEB space density is (1.4\,$^{+0.8}_{-0.6}$)\,$\times$\,10$^{-4}$\,pc$^{-3}$. The small number statistics of the PCEBs in the 20\,pc volume generate large error bars on the space density. \citet{Schreiber2003} estimated a lower limit on the PCEB space density of 1\,$\times$\,10$^{-5}$\,pc$^{-3}$, and more recently \citet{Shariat2026} used a sample of eclipsing short-period WD\,$+$\,dM PCEB systems in 200\,pc to determine a space density of 7.2\,$\times$\,10$^{-5}$\,pc$^{-3}$ for systems with orbital periods less than 2\,days. We cannot directly compare our results with those of \citet{Shariat2026} because our sample considers all periods (the period of \gtwo\ is greater than 2\,days), and is so small that it may not be representative of the wider WD\,$+$\,dM PCEB population. 

In the \gaia\ Catalogue of Nearby Stars \citep{GCNS2021}, which is assumed to be almost complete out to 20\,pc, there are 2007 M\,dwarfs. We use this number to estimate the fraction of M\,dwarfs that have been surveyed for RV measurements, enabling a (non-eclipsing) PCEB system to be identified. Between 2003 and 2019, the HARPS instrument obtained radial velocities of 316 M\,dwarfs in the southern hemisphere within 20\,pc \citep{Mignon2024}. The CARMENES survey has obtained radial velocities of 308 M\,dwarfs in the northern hemisphere \citep{Quirrenbach2018,Ribas2023}. Assuming minimal overlap, these two surveys alone provide RV coverage for at least 30\,per\,cent of the 20\,pc M\,dwarf population. However, this represents an upper limit for the fraction of stars with sufficient data for PCEB identification, as many of these targets lack enough epochs to confirm a companion. Other smaller surveys including the HARPS--N HADES survey \citep{Perger2017} cover smaller subsets of M\,dwarfs that overlap with those in the CARMENES and HARPS programmes. For improvements in PCEB statistics, a substantially larger fraction of M\,dwarfs require multi-epoch RV monitoring. A crude estimate of uniform detectability of PCEBs implies that up to 9\,--\,10 more WD\,$+$\,dM PCEBs with M dwarf companions could be identified if the remaining 70\,per\,cent of M\,dwarfs were surveyed with RV instruments. \gaia\ DR4 will soon provide full coverage of epoch astrometry for local M\,dwarfs, increasing the number of identified astrometric WD\,$+$\,dM binaries. 

\subsection{BPASS predictions for PCEB population}

To assess whether the number of PCEBs we have observed within 20\,pc is consistent with predictions from population synthesis, we compared our results with Binary Population and Spectral Synthesis (\textsc{bpass}) grids \citep{Stanway2018,Stevance2020,Byrne2022}. \textsc{bpass} models explicitly account for binary star systems, making them ideal for our use-case. We chose a \textsc{bpass} grid with Solar metallicity and constant star formation history, both of which are consistent with the local Solar neighbourhood \citep{Cukanovaite2023}. Within this grid, we filtered all outputs that are currently PCEBs at the age of the Galactic disc (10\,Gyr). Our criteria for a PCEB system was: no current Roche lobe overflow, an orbital period of 20\,days or less, and one (but not both) of the stars in the binary is a white dwarf. 

The \textsc{bpass} grids contain models that have binary stars labelled as Star 1 and Star 2. Models of Type 1 consider Star 1 to be the more massive of the binary, and therefore Star 1 evolves into a white dwarf first. For this model type, we checked that Star 1 was a white dwarf, and that Star 2 was not a white dwarf. For Star 1 to be a white dwarf, its radius must be smaller than 0.03\,\Rsun, the mass of hydrogen must be $<$\,10$^{-3}$\,\Msun, and the core must be degenerate, either with a CO or ONe core or with more than 0.3\,\Msun\ in a He core for a low-mass white dwarf. For Star 2 to be a main sequence star, its radius must be greater than 0.1\,\Rsun. Models of Type 2 consider a scenario where Star 2 has reached the end of its evolution, and Star 1 is still evolving. For this model type, we checked that Star 1 was still a main sequence star by checking it had a radius greater than 0.1\,\Rsun. The criteria for main sequence stars includes all spectral types earlier than, and including, M\,dwarfs. However, the output is significantly dominated by M\,dwarfs, due to local demographics.

We applied the above cuts to the \textsc{bpass} grids and normalised to 20\,pc using the volume calculation from Section~\ref{sec:spacedensity}, compared to the total number of white dwarfs of any kind, single or in a binary, in the \textsc{bpass} grid, and scaled to the observed number of white dwarfs in 20\,pc (153). We recovered 4.4 PCEB systems within the local 20\,pc volume from the \textsc{bpass} grid. The result of our calculation agrees well with the four observed PCEBs in 20\,pc discussed in this work. There are some known PCEBs that undergo a partial CE phase, however we did not consider these in our \textsc{bpass} calculation, as in the 20\,pc volume we have only detected short-period PCEBs, and the longer-period systems are a somewhat distinct and therefore separate population. The four direct detections of PCEBs within 20\,pc are clearly small-number statistics, and should not be used to infer implications on the overall Galactic binary population. 

The \textsc{bpass} output is qualitatively different to that of \citet{Toonen2017} who, using an independent population synthesis code \textsc{SeBa}, predicted that 0.5\,--\,1\,per\,cent of white dwarfs are part of unresolved WD\,$+$\,MS systems in the local Solar neighbourhood. The four PCEBs within 20\,pc are unresolved with \gaia\ and as such we consider these quantities interchangeable. Given the number of confirmed white dwarfs within 20\,pc (153), this fraction corresponds to 0.7\,--\,1.5 white dwarfs from \citet{Toonen2017}. With \textsc{bpass}, the post-CE separation and stellar masses are based on best efforts to model the primary star’s response to the CE phase, rather than fixed at a set $\alpha$\textsubscript{CE} efficiency parameter. \citet{Toonen2017} used fixed values of $\alpha$\textsubscript{CE}, which may have caused the difference in PCEB fractions between the two codes.

\begin{figure}
    \centering
    \includegraphics[width=\columnwidth]{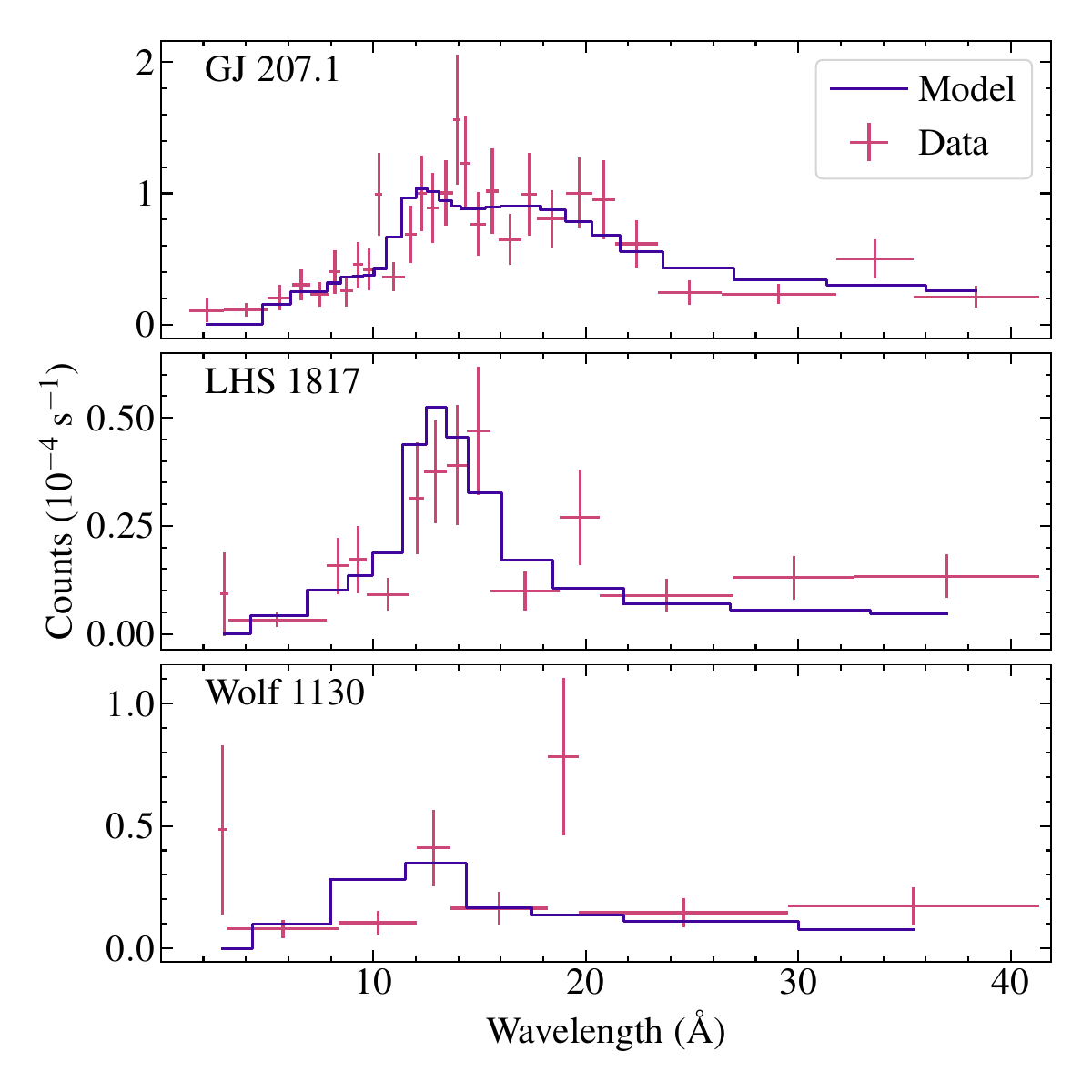}
    \caption{XRT spectra for the three detected systems, along with the APEC model fits used to measure the X-ray fluxes.}
    \label{fig:xrtspecs}
\end{figure}

\subsection{X-ray fluxes}
\textit{Swift} XRT allows us to compare the activity of the rapidly rotating, spun up M\,dwarfs in our binaries with the larger population of single M\,dwarfs. XRT light curves and spectra for each \textit{Swift} visit were extracted using the online XRT tool\footnote{\url{https://www.swift.ac.uk/user_objects/}}. The light curves were used to select which of the \textit{Swift} spectra to fit. For \gj\ and \lhs\, the light curves were flat. For \gj\ we used only the spectrum from the simultaneous observation with \hst\ as it had sufficient \snr. For \lhs\ the \snr\ was lower so we combined all available observations into one spectrum. The light curve of \wolf\ showed clear signs of a flare in the first observation, although unfortunately the peak of the flare occurred during an occultation by Earth. We therefore used only the second observation. \gtwo\ was not detected in the combined \textit{Swift} image with a 2000\,s exposure time.

Assuming that the white dwarfs do not contribute to the X-ray flux, we fitted the spectra of the three detected stars using XSPEC with APEC plasma models with one (\lhs, \wolf) or two (\gj) temperatures, modified by a fixed hydrogen column density of $\log$\,N(\ion{H}{i}) $=18$. Coronal abundances were fixed to 0.4\,Solar. From the models we estimated X-ray fluxes in the ROSAT band (0.1--2.4\,keV). We calculated an upper limit on the count rate for \gtwo\ of $5.5\times10^{-3}$ counts\,s$^{-1}$ using the {\sc HEASOFT sosta} routine and converted it into a flux with WebPIMMS using a 0.5\,keV APEC model. Given the poor \snr\ of the \wolf\ XRT spectrum, we double checked our result by converting the average count rate of the observation into a flux in the same manner, finding an X-ray flux consistent with the spectroscopic result. XRT spectra for the three detected systems are shown in Fig.~\ref{fig:xrtspecs}, and X-ray luminosities for all four stars are given in Table~\ref{tab:fitting_params}. 

\subsection{Orbital and rotation period of G203-47}

\begin{figure}
    \centering
	\includegraphics[width=\columnwidth]{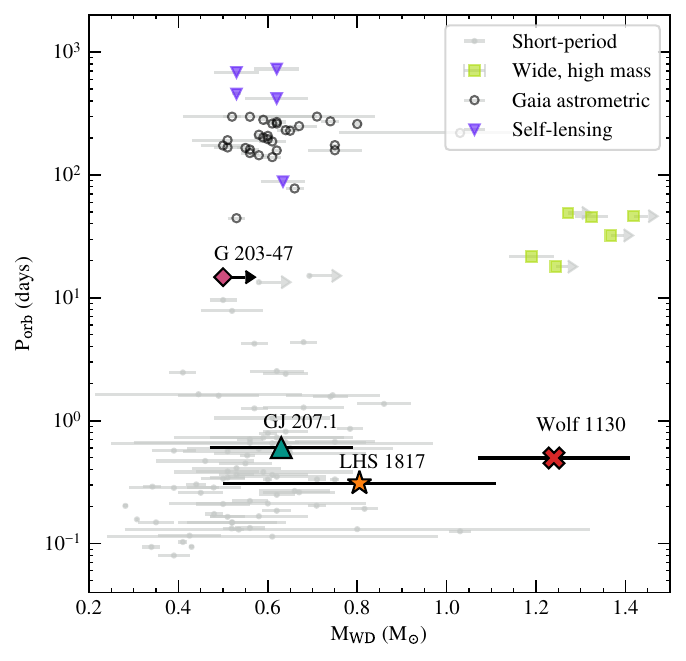}
	\caption{The orbital period (P\textsubscript{orb}) against white dwarf mass (M\textsubscript{WD}) of PCEBs, adapted from Fig. 7 of \citet{Yamaguchi2024c}. The four PCEBs discussed in this work are shown by the following symbols: \gtwo\ is the pink diamond, \gj\ is the teal triangle, \lhs\ is the orange star, and \wolf\ is the red cross. The data for the other systems are from the following: short-period PCEBs are from \citet{Zorotovic2010,Rebassa2012,Hernandez2021,Hernandez2022a,Hernandez2022b,Lin2026,Cavalier2026}, wide PCEBs are from \citet{Wonnacott1993,Corcoran2021,Yamaguchi2024a}, \gaia\ astrometric PCEBs are from \citet{Yamaguchi2024c}, and self-lensing PCEBs are from \citet{Kruse2014,Kawahara2018,Yamaguchi2024b}.}
    \label{fig:g203_47_yamaguchi}
\end{figure}

Given the 14.9-day orbital period of \gtwo, and its M\,dwarf mass determined in Table~\ref{tab:fitting_params}, it sits in a unique region on the orbital period against white dwarf mass parameter space (see Fig.~\ref{fig:g203_47_yamaguchi}), which is an adapted version of Fig. 7 from \citet{Yamaguchi2024c}. The two main PCEB populations, those with longer orbits and those with shorter orbits, are thought to be the result of two different formation channels. The white dwarf progenitors in the short period PCEBs started closer to their companions and were fully engulfed in the envelope \citep{Zorotovic2010}. The wider orbit PCEBs underwent only brief CE interactions in the Asymptotic Giant Branch \citep{Yamaguchi2024c}, or could have also formed through stable mass transfer \citep{Lagos2022}. Given that \citet{Delfosse1999} only provided a lower limit for the mass of the white dwarf in \gtwo, and the white dwarf mass could not be constrained from our STIS fit, it could possibly sit in the region populated by the high-mass wide PCEBs identified by \citet{Yamaguchi2024a}. However, the population of PCEBs with longer orbital periods also have more massive companions compared to the M\,dwarf in \gtwo.

\begin{figure}
    \centering
    \includegraphics[width=\columnwidth]{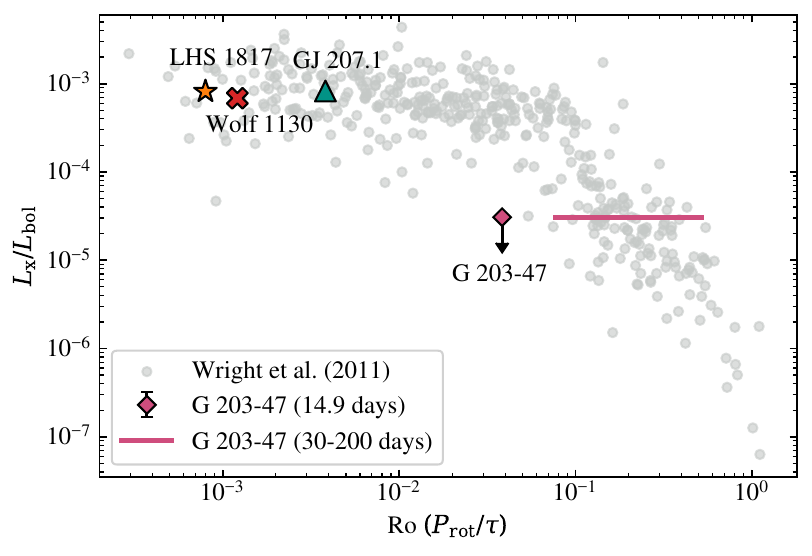}
    \caption{Fractional luminosities as a function of Rossby number for the M\,dwarf components of our PCEBs and the sample of single M\,dwarfs from \citet{Wright2011}. Rossby numbers (Ro) were computed using MESA-based convective turnover times ($\tau$). The four PCEBs discussed in this work are shown by the following symbols: \gj\ is the teal triangle, \lhs\ is the orange star, \wolf\ is the red cross, and for \gtwo\ the pink diamond shows the assumption that the system is tidally locked, whereas the pink bar assumes the system is not tidally locked.}
    \label{fig:lxros}
\end{figure}

Figure~\ref{fig:lxros} shows the fractional X-ray luminosities as a function of Rossby number (Ro $=$ rotation period divided by the convective turnover time), compared with a sample of stars from \citet{Wright2011}. \citet{Barraza2026} found that commonly used prescriptions for the convective turnover time, such as those derived by \citet{Wright2011} and \citet{Wright2018}, can differ significantly from more detailed Modules for Experiments in Stellar Astrophysics (MESA) calculations. In particular, their Fig.~1 compares the MESA-based values with the \citet{Wright2011} derivation. Therefore in Fig.~\ref{fig:lxros}, we show the MESA-calculated convective turnover times. We calculated the Rossby numbers for our PCEBs using the luminosities in Table~\ref{tab:fitting_params}, and obtained average convective turnover times using the MESA models. If we assume that \gtwo\ is tidally locked, i.e. its rotation period is equal to its orbital period, we find it does not sit within the main sample in Fig.~\ref{fig:lxros}. Therefore, it is possible that \gtwo\ is not tidally locked. 

To test the hypothesis that \gtwo\ is not tidally locked, we searched available photometric data. At periods longer than a couple of days, the contributions to a PCEB light curve from effects tied directly to the orbital period (e.g. ellipsoidal modulation and reflection) become small, and rotational modulation due to surface features on the M\,dwarf (spots, faculae) dominate. Photometric variation therefore provides an independent measurement of the M\,dwarf rotation period. Unfortunately the 14.9-day orbital period of \gtwo\ is close to the \textit{TESS} orbital period (13.7\,days), and the \textit{TESS} light curve is dominated by systematics on longer time scales. The best available light curve comes from the MEarth survey, from which we find a tentative 126.3-day period using the analysis techniques described in \citet{Newton2016}. The quiescent H$\alpha$ flux measured by \citet{Newton2017} also implies a rotation period $>$\,100\,days. While none of these measurements provide an exact rotation period for the \gtwo\ M\,dwarf, together they strongly imply that the rotation period is much longer than the orbital period and that the system is not tidally locked. Fig.~\ref{fig:lxros} shows that between a rotation period of 30 and 200 days, \gtwo\ lies well within the unsaturated regime. However, since the X-ray measurement may be much lower, this range can change significantly. 

There are only two PCEBs known to have orbital periods close to \gtwo: 2MASS\,J14544500$+$4626456 (hereafter 2M\,1454) and 2MASS\,J04472847$+$2444523 (hereafter 2M\,0447). Both are shown in Fig.~\ref{fig:g203_47_yamaguchi} as the grey datapoints to the right of \gtwo\ with horizontal lower limits. \citet{Corcoran2021} identified a $\approx$\,15.1\,day orbital period from radial velocity data for 2M\,1454. Using Zwicky Transient Facility data, we confirm this period and find that, unlike \gtwo, the main-sequence component either is tidally locked to the orbital period or is very close to it. \citet{Lin2026} identified an orbital period of 13.4\,days from \gaia\ RVs for 2M\,0447, and they show that it is most likely a white dwarf plus K\,dwarf PCEB. The \textit{Kepler} light curve of 2M\,0447 shows a clear $\approx$\,10\,day period which is likely the rotation period of the main-sequence star; several days faster than the orbital period (Appendix \ref{sec:ztf}, Figure \ref{fig:ztf}). 

We find no clear cut off in orbital period between systems that are and are not tidally locked. Furthermore, the cooling age of the white dwarf in 2M\,1454 is $\approx$\,0.3\,Gyr, compared with $\approx$\,4\,Gyr for \gtwo\ (the white dwarf temperature of 2M\,0447 and thus the cooling age cannot be measured with existing data). If the two systems have undergone the same formation process then we would expect \gtwo\ to be closer to tidal locking simply because it has had more time. Instead, the rotation periods suggest that the three systems may have different histories. The orbital periods are too long for them to have tidally locked in the PCEB phase \citep{Hilditch2001}, so we speculate that the key event is the CE phase.  

As shown in Figure \ref{fig:g203_47_yamaguchi}, the known PCEBs appear to be split into two populations, with \gtwo, 2M\,1454, and 2M\,0447 falling in the middle. The different tidal histories suggest that 2M\,1454, 2M\,0447 and \gtwo\ may represent extreme cases at the boundaries of the two populations. In this scenario, \gtwo\ and  2M\,0447 are on the extreme low end of the long-period population, undergoing only a relatively short CE event in the AGB phase and not transferring enough angular momentum to be tidally locked. Conversely, 2M\,1454 may represent the extreme high end of the short-period population, undergoing a classical, long CE event in the RGB phase and ending up tidally locked to its companion. This hypothesis is supported by the detection of s-process elements in the main sequence component of 2M\,0447, indicating that the white dwarf progenitor did indeed go through the AGB \citep{Lin2026}. Efforts to further test this scenario would benefit from modelling of angular momentum transfer and rotation period evolution during the CE phase, as well as  main-sequence star rotation period measurements for the handful of known systems with several-day orbital periods. Whilst their distances and faintness make X-ray observations impractical and photometric measurements challenging, H\,$\alpha$ spectroscopy may be sufficient to at least approximate their rotation periods following \citet{Newton2017}. 

\section{Conclusions}
\label{sec:conclusion}

We have directly observed, for the first time, the white dwarf components of four PCEB systems within 20\,pc of the Sun: \gtwo, \gj, \lhs\ and \wolf. These systems were identified as PCEBs following radial velocity measurements of their M\,dwarfs indicating a degenerate companion, and we followed them up with \hst\ STIS using the near-UV G230L and G230LB gratings. We applied a custom correction function to the G230LB data which suffers from red leak, in order to improve the flux calibration of the spectra. The spectra showed many emission features due to activity, which were masked to isolate the white dwarf continuum prior to fitting, which highlights the necessity of near-UV spectra rather than broad-band photometry for accurate white dwarf fitting. We fitted the STIS spectra with a combination of white dwarf models and M\,dwarf proxy spectra in order to estimate the effective temperatures of the white dwarf components of the PCEB systems, and identified white dwarfs in all four systems with temperatures ranging from 5300\,K to 6300\,K. These four white dwarfs are some of our closest stellar neighbours, and in particular the white dwarf in \gtwo\ is the ninth closest white dwarf to the Sun.

We used these four new white dwarfs to update the local white dwarf space density using the 20\,pc volume: (5.2\,$\pm\,$0.4)\,$\times$\,10$^{-3}$\,pc$^{-3}$. We also determined the local PCEB space density: (1.4\,$^{+0.8}_{-0.6}$)\,$\times$\,10$^{-4}$\,pc$^{-3}$. We highlighted the need for more M\,dwarfs to be surveyed for radial velocity measurements, in order to identify more local PCEB systems, and estimated that up to 9\,--\,10 new PCEBs could be identified within 20\,pc if all M\,dwarfs were surveyed. \textsc{bpass} grids predicted 4.4 PCEBs within 20\,pc, which is in good agreement with the four we have observed. 

Additionally, we studied the rotation and orbital motion of \gtwo, which sits in an unusual position in-between the two distinct PCEB populations in orbital period space. We found that it has a 14.9-day orbital period but a rotation period possibly $>$\,100\,days, making the system not tidally locked. We compared \gtwo\ to two other PCEBs with similar orbital periods, 2MASS\,J14544500$+$4626456 and 2MASS\,J04472847$+$2444523, with the former found to be tidally locked and the latter not. These systems are at the boundaries of the two PCEB populations, with \gtwo\ likely being part of the long-period population which have only undergone a partial CE phase.

\section*{Acknowledgements}

MOB, PET and BTG received funding from the European Research Council under the European Union’s Horizon 2020 research and innovation programme numbers 101002408 (MOS100PC; MOB and PET) and 101020057 (WDPLANETS; BTG). FLV acknowledges support from ANID FONDECYT Postdoctoral Grant No. 3250464. JABJ acknowledges support from ANID Becas/Doctorado Nacional 21263186. Support for \hst\ PID \pid\ was provided by NASA through a grant from the Space Telescope Science Institute.

This work was based on observations made with the NASA/ESA Hubble Space Telescope, obtained from the Data Archive at the Space Telescope Science Institute, which is operated by the Association of Universities for Research in Astronomy, Inc., under NASA contract NAS 5--26555. These observations are associated with PID 9786 and PID \pid, and we thank Carol Rodriguez and Emily Rickman for their assistance with PID \pid. All of the \hst\ data presented in this paper were obtained from the Mikulski Archive for Space Telescopes (MAST). We thank the \textit{Swift} operations team for enabling the coordinated observations with \hst. We acknowledge the use of public data from the \textit{Swift} data archive. This work made use of v2.3 of the Binary Population and Spectral Synthesis (\textsc{bpass}) models as last described in \citet{Byrne2022} and \citet{Stanway2018}. We thank Elisabeth Newton for the analysis of the MEarth data for \gtwo. We thank the referee for their helpful comments.

\section*{Data Availability}
 
The \hst\ data underlying this article are available in the Barbara A. Mikulski Archive for Space Telescopes (MAST) at: \url{https://mast.stsci.edu/portal/Mashup/Clients/Mast/Portal.html}, and can be accessed by searching for PID 17778. The \gtwo\ data are subject to an embargo of 6 months from the observation date, which was 2026 March 25. Once the embargo expires this data will also be available in the MAST portal. The \textit{Swift} data are available in the HEASARC Archive at: \url{https://heasarc.gsfc.nasa.gov/cgi-bin/W3Browse/swift.pl}.



\bibliographystyle{mnras}
\bibliography{mybib} 



\newpage
\appendix

\section{Photometric periods of 2MASS\,J14544500+4626456 and 2MASS\,J04472847$+$2444523}
\label{sec:ztf}
\begin{figure*}
    \centering
    \includegraphics[width=\textwidth]{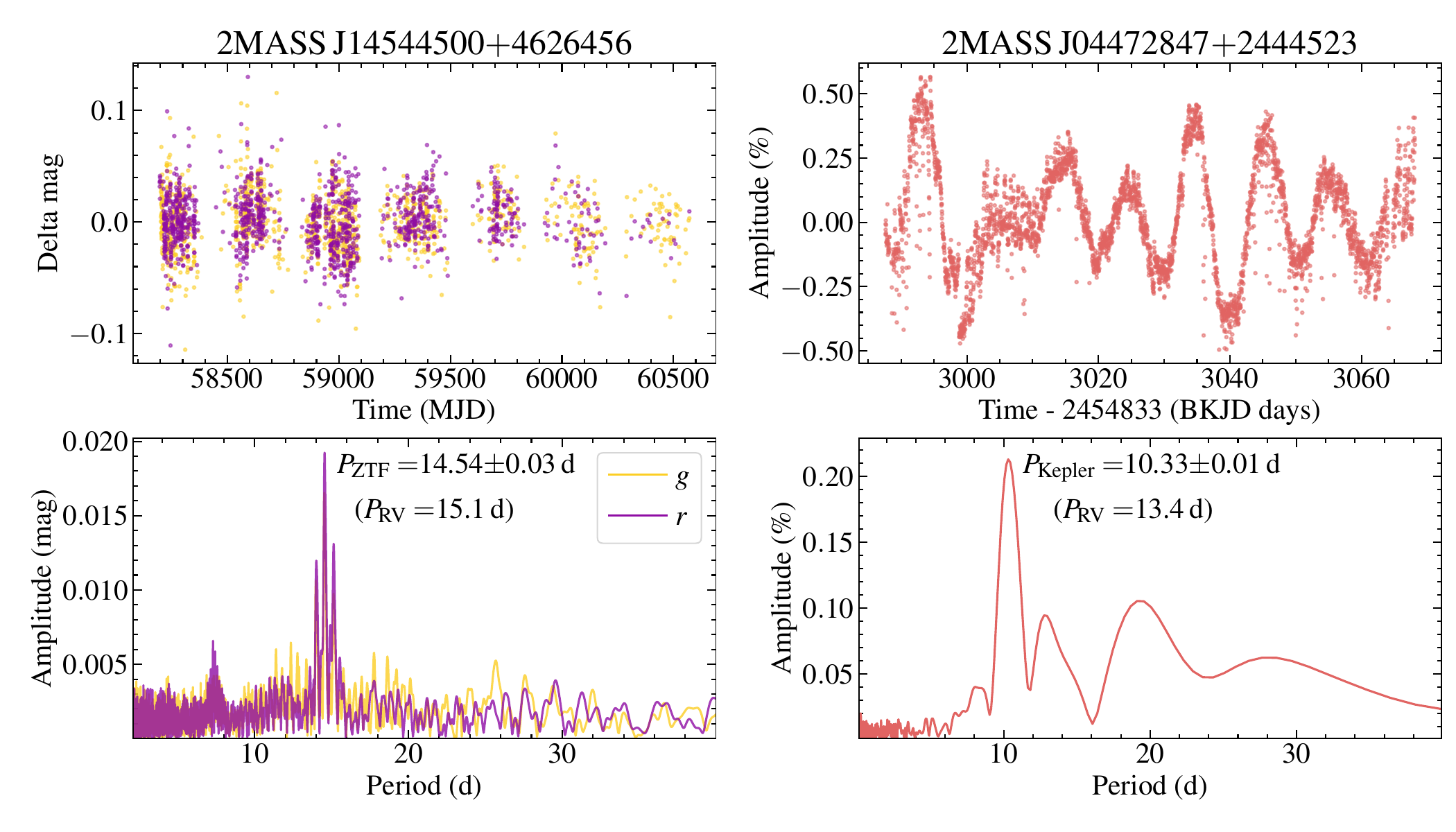}
    \caption{Photometric data from ZTF (left) and Kepler (right) for two intermediate PCEBs. The bottom panels show the Lomb-Scargle periodograms computed from the light curves.}
    \label{fig:ztf}
\end{figure*}

To support the analysis of \gtwo\ we have measured the rotation periods of the main-sequence stars in two PCEBs with similar $\approx$15\,d periods using archival light curves. \citet{Corcoran2021} identify an $\approx$15.1\,d orbital period for 2MASS\,J14544500+4626456 from seven epochs of APOGEE spectroscopy. We retrieved $g-$ and $r-$band ZTF photometry from ZTF Data Release 23 \citep{Masci2019}, finding 1324 and 1241 measurements respectively obtained over 6.5 years. The left panels of Figure \ref{fig:ztf} show the light curves and periodogram, demonstrating a clear periodic signal that is almost certainly due to spot modulation on the main-sequence component of the system. We fit the light curves separately with a sine wave based on the peak of the periodogram and find an average photometric period of $14.54\pm0.03$\,d, both confirming the approximate period from \citet{Corcoran2021} and that the main-sequence component is, or is very close to, tidally locked to the orbital period.   

For 2MASS\,J04472847$+$2444523, \citet{Lin2026} find a radial velocity period $P_{\mathrm{orb}} = 13.3968\pm0.00058$\,d. The target was observed by K2 \citep{Howell2014} at 30\,minute cadence almost continuously for 80\,days. The light curve shows spot modulation with a regular period but highly irregular amplitude. We measured the period as before and find $P_{\mathrm{Kepler}}= 10.33\pm0.01\,d$, somewhat faster than the period measured from radial velocities. 

We additionally used the VOSA tool \citep{Bayo2008} to find an approximate temperature for the 2MASS\,J14544500+4626456 white dwarf and thus the cooling age. Fitting a combination of a \textsc{BT-Settl-CIFIST} M\,dwarf model and Koester white dwarf model with fixed \logg\,$=$\,8 \citep{Koester2010}, we find a white dwarf \Teff=12750\,K, corresponding to a cooling age of $\approx0.3$\,Gyr. As a more detailed analysis of the data is beyond the scope of this paper we do not present this a formal measurement of the white dwarf \Teff, but it is sufficient to show that the cooling age is much less than \gtwo. Unfortunately no ultraviolet data is available for 2MASS\,J04472847$+$2444523 so we cannot estimate the white dwarf temperature.    

\bsp	
\label{lastpage}
\end{document}